# Spin ice in a field: quasi-phases and pseudo-transitions.

P.N. Timonin
Southern Federal University, Rostov-on-Don, 344090, Russia.

*Thermodynamics of the short-range model of spin ice magnets in a field is considered in the Bethe – Peierls approximation. The results obtained for [111], [100] and [011] fields agrees reasonably well with the existing Monte-Carlo simulations and some experiments. Quite remarkably all extremely sharp field-induced anomalies are described in this approximation by the analytical functions of temperature and applied field. In spite of the absence of true phase transitions the analysis of the entropy and specific heat reliefs over H-T plane allows in most cases to discern the "pseudo-phases" with specific character of spin fluctuations and outline the curves of more or less sharp "pseudo-transitions" between them.*

The discovery of spin ice compounds [1, 2] has opened a wide perspective in the studies of real geometrically frustrated magnets with their reach physics stemming from the macroscopically degenerate ground states. The more so as they can be described by the relatively simple Ising model with the nearest-neighbour exchange on the pyrochlore lattice. This is due to the lucky chance that strong dipole interactions in these compounds have a negligible effect on the low-energy excitations of the Ising moments directed along the lines connecting the centres of corner-sharing tetrahedra [3]. So low-temperature physics of spin-ices can be adequately captured by the short-range Ising model except for the ultra low temperatures where the equilibrium properties may be unobservable [4].

Such model predicts the absence of phase transitions in zero field in accordance with experiments in the (established) acknowledged spin ice compounds [1, 2]. Meanwhile a wealth of more or less sharp anomalies in the applied magnetic fields $H$ of different directions is observed in their thermodynamic parameters [5-14]. Some of these anomalies are interpreted as the field-induced transitions while others are thought to indicate the crossover between the regions with different types of collective spin fluctuations. The notion of such regions originates from the Villain's idea of low-temperature "spin-liquid" state in frustrated magnets [15] where spin fluctuations are strongly correlated being confined mostly to the ground states' subspace. In contrast the high temperature region features uncorrelated spin fluctuations thus representing a genuine paramagnet. While there is no true phase transition between paramagnet and spin-liquid state still the temperature dividing these "quasi-phases" can be pointed out – it is the temperature $T_m$ of specific heat maximum in its temperature dependence $C(T)$ [4]. Indeed, this maximum indicates the more or less sharp drop of entropy due to the confinement of spin fluctuations at low $T$. One may hope that such definition of $T_m$ can justify the notion of "pseudo-transition" between the "quasi-phases" with different types of spin fluctuations and may help to quantify in the framework of rigorous theory the regions where various spin-liquid states exist.

Implicitly the notions of "quasi-phases" and "pseudo-transition" are widely used to interpret heuristically the observed field-induced anomalies of $C(T)$ in spin ices and to identify the regions belonging to the different spin-liquid states on the $H$-$T$ planes [9-13]. Yet it is important to discriminate the "pseudo-transitions" and the ordinary ones as the microscopic models describing first of them would not have any singular point but only the crossover regions. The more so as these crossovers can grow more and more sharp at low $T$ and in the vicinity of critical fields so the "pseudo-transitions" may look like true ones in experiments and simulations. Thus the observed sharpening of "pseudo-transitions" give rise to the idea that spin-ice compounds in [111] [9-11], [110] [12,13] and [100] [16] fields experience at low temperatures the first-order transitions of "gas-liquid" type ending up at the critical point at some maximal $T$. Also from the same sharpening the notion of the specific "Kasteleyn transition" in the short-range spin-ice model in [100] field [17] originated.

Here we show that most probably there are no true phase transitions in the short-range spin-ice model in [111], [100] and [110] fields and all observed thermodynamic anomalies can be described by the perfectly smooth functions of $T$ and $H$. Actually this conclusion can be made on the basis of the existing theoretical results. Indeed, the papers presenting the Monte-Carlo simulations in the regions of "gas-liquid" [16, 18] and "Kasteleyn" transitions [17] present also the results obtained in the Bethe-Peierls (BP) approximation "corroborating" the numerical ones. Both these data show a remarkable agreement with experiments [6, 9-11] but BP results describe the extremely sharp anomalies with perfectly smooth functions. Due attention to this fact may tell us that actually we deal with the "pseudo-transitions" and their sharpening at low *T*.

Here we show explicitly how the BP approximation in the short-range spin-ice model manages to describe very sharp field-induced anomalies with analytical functions of *T* and *H*. In the BP approach the sharp anomalies at low *T* are the reflection of true first-order transitions which take place at *T*=0 and some critical fields. Noticing the high precision with which BP approximation can reproduce the nominally exact results [16-18] as well as experimental data [6-13] and quite clear physics underlying the origin of appearing anomalies we can be sure that the BP's "pseudo-transitions" are not artefacts of the approximation but rather the intrinsic feature of the model.

# 1. Spin ice in Bethe-Peierls approximation

Magnetic ions in spin ice are placed on the pyrochlore lattice consisting of corner-sharing tetrahedra the fragment of which is shown in Fig. 1. Strong anisotropy allows only two directions of magnetic moments along the local easy axis connecting the site with the centres of tetrahedra. So the magnetic state is defined by Ising spins $\sigma_\alpha$ on the sites. Considering four sites belonging to the central tetrahedron in Fig.1 we define their easy axes by the unit vectors shown in this figure

$$\mathbf{e}_0 = (\hat{x}+\hat{y}+\hat{z})/\sqrt{3}, \quad \mathbf{e}_1 = (\hat{x}-\hat{y}-\hat{z})/\sqrt{3}$$
$$\mathbf{e}_2 = (-\hat{x}+\hat{y}-\hat{z})/\sqrt{3}, \quad \mathbf{e}_3 = (-\hat{x}-\hat{y}+\hat{z})/\sqrt{3} \tag{1}$$

Here $\hat{x}, \hat{y}, \hat{z}$ are the unit vectors along the coordinate axes.

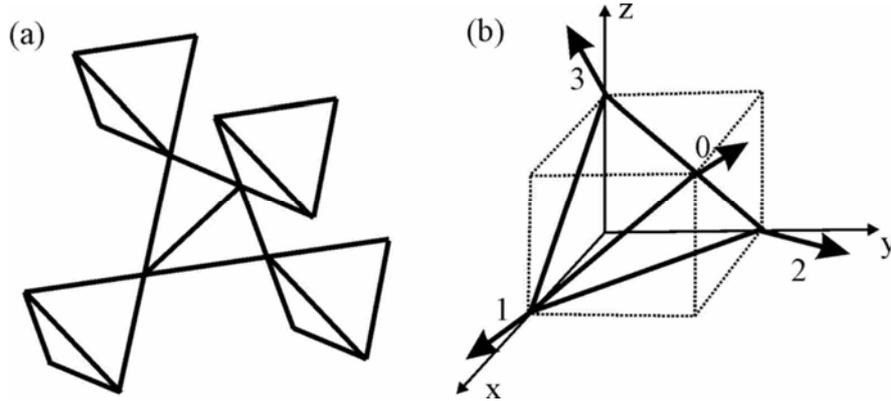

Fig.1. Fragment of pyrochlore lattice (a) and vectors defining the easy axes directions on their sites (b).

Thus the magnetic moments of the sites are
$$\mathbf{m}_\alpha = \mathbf{e}_\alpha \sigma_\alpha \tag{2}$$
and the effective Hamiltonian for the tetrahedron is [14]
$$\mathcal{H}(\sigma) = \frac{J}{2}\left(\sum_{\alpha=0}^{3}\sigma_\alpha\right)^2 - \sum_{\alpha=0}^{3}\sigma_\alpha H_\alpha, \quad H_\alpha \equiv \mathbf{H}\mathbf{e}_\alpha \tag{3}$$
With the definition (1) the following identities hold
$$\sum_{\alpha=0}^{3}\mathbf{e}_\alpha = 0, \quad \mathbf{e}_\alpha\mathbf{e}_\beta = -\frac{1}{3}, \quad \alpha \neq \beta \tag{4}$$
and $\sigma_\alpha = 1$ corresponds to the "out" moment.

The BP approximation for the pyrochlore lattice [17] implies that the effective fields acting on the sites of the given tetrahedron (say, the central one in Fig. 1) from all other sites are the same as those acting on its nearest neighbours (outer sites in Fig. 1) from all other sites except those of a given tetrahedron. This field equivalence does not really hold on the pyrochlore lattice due to the correlations arising from the closed loops of tetrahedra. But it becomes exact on the variant of the hierarchical Bethe lattice built from the corner-sharing tetrahedra [17]. We can get it from the cluster in Fig. 1 attaching to the each outer site a tetrahedron then endowing each appeared new outer site with a new tetrahedron and so on. The process is illustrated by the plane graph in Fig. 2a, where tetrahedra are projected to the squares with the numbers on their sites indicating the easy axes orientations.

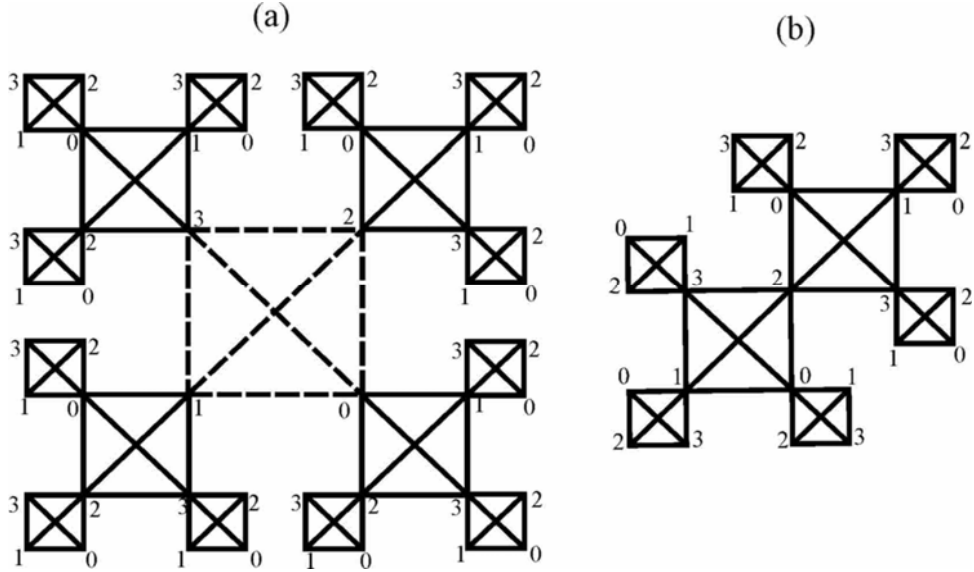

Fig.2. (a) Fragment of Bethe lattice built hierarchically via addition of new tetrahedra to the outer sites. It can be viewed as consisting of four trees connected by the bonds of the central tetrahedron (dashed lines). The numbers indicate the directions of easy axes of the sites. (b) The same lattice can be obtained from two trees via merging their root sites.

As Fig. 2a shows we can consider such Bethe lattice as consisting of four $\alpha$-trees with $\alpha = 0$, 1, 2, 3 denoting their root sites which are connected by the bonds of the central tetrahedra. Otherwise we can get it from two $\alpha$-trees merging their root sites, cf. Fig.2b. So there are two ways to construct the free energy of Bethe lattice with the partial partition functions $Z_\alpha(\sigma_\alpha, N)$ for the N-site $\alpha$-trees summed over all spins except the root one. Via the first way we have the free energy

$$\Phi_4 = -T \ln Tr_\sigma e^{-\beta \mathcal{H}(\sigma)} \prod_{\alpha=0}^{3} Z_\alpha(\sigma_\alpha, N) \tag{5}$$

where $\beta = T^{-1}$ and $Tr_\sigma$ denotes the summation over $\sigma_\alpha$, $\alpha = 0, 1, 2, 3$.
Via the other way we get the free energies

$$\Phi_{2\alpha} = -T \ln Tr_{\sigma_\alpha} e^{h_\alpha \sigma_\alpha} Z_\alpha^2(\sigma_\alpha, N), \quad h_\alpha \equiv \beta H_\alpha. \tag{6}$$

Assuming the finite correlation range in our system we have in the thermodynamic limit, $N \to \infty$,

$$\Phi_4 \to 4NF + 4SG, \quad \Phi_{2\alpha} \to (2N-1)F + 2SG$$

where $F$ is the (internal) free energy per site in the this limit, $S$ is the number of surface sites in a tree and $G$ is the density of the surface contribution to the full potential. 2N-1 in the second relation appears because we have merged two root sites into one.

The peculiarity of the Bethe lattices having formally the infinite spatial dimension consists in the finite ratio $S/N$ at $N \to \infty$. In our lattice $S/N \to 2/3$ and we cannot neglect the surface contribution to the full potential. Yet this circumstance does not prevent the determination of the free energy per site $F$ from $\Phi_4$ and $\Phi_{2\alpha}$ as we can choose their linear combination where the surface terms conceal each other. Thus we have

$$F = \frac{1}{4} \lim_{N \to \infty} \left( 2\Phi_4 - \sum_{\alpha=0}^{3} \Phi_{2\alpha} \right) \tag{7}$$

We can also represent $Z_\alpha(\sigma_\alpha, N)$ in Eqs. (5, 6) as

$$Z_\alpha(\sigma_\alpha, N) = A_\alpha(N) e^{x_\alpha \sigma_\alpha}$$

Here $x_\alpha T$ have the meaning of the effective magnetic fields exerted on the sites of the central tetrahedron by the other spins in the lattice. Then introducing instead of $x_\alpha$ new variables

$$f_\alpha = x_\alpha + h_\alpha$$

we get from (3), (5-7)

$$F = \frac{T}{4}\sum_{\alpha=0}^{3}\ln\left[2\cosh(2f_\alpha - h_\alpha)\right] - \frac{T}{2}\ln 2Z(\mathbf{f}), \quad Z(\mathbf{f}) \equiv \frac{1}{2}Tr_\sigma w(\sigma,\mathbf{f}) \tag{8}$$

$$w(\sigma,\mathbf{f}) = \exp\left[-\frac{K}{2}\left(\sum_{\alpha=0}^{3}\sigma_\alpha\right)^2 + \sum_{\alpha=0}^{3}f_\alpha \sigma_\alpha\right], \quad K \equiv \beta J \tag{9}$$

Owing to the form of Eq. (7) the resulting expression for $F$ depends only on the parameters $f_\alpha$. The equations for $f_\alpha$ can be obtained from the condition that they provide a minimum of $F$,

$$\frac{\partial F}{\partial f_\alpha} = 0, \quad \sum_{\alpha,\beta} c_\alpha c_\beta \frac{\partial^2 F}{\partial f_\alpha \partial f_\beta} > 0 \text{ for all } \mathbf{c}, \tag{10}$$

or

$$\tanh(2f_\alpha - h_\alpha) = \langle \sigma_\alpha \rangle_w, \tag{11}$$

where the angular brackets with the "w" subscript denote the average with the distribution function $w(\sigma,\mathbf{f})$ from Eq. (9). Otherwise we can get the same equations (11) considering the recursion relations for $Z_\alpha(\sigma_\alpha, N)$. For the Gessian in Eq. (10) we get

$$G_{\alpha\beta} \equiv \frac{\partial^2 F}{\partial f_\alpha \partial f_\beta} = \delta_{\alpha\beta}\left[1 - \langle \sigma_\alpha \rangle_w^2\right] - \frac{1}{2}\langle \sigma_\alpha \sigma_\beta \rangle_w + \frac{1}{2}\langle \sigma_\alpha \rangle_w \langle \sigma_\beta \rangle_w \tag{12}$$

Actually the positive definiteness of Gessian does not arise in the standard derivation of Eqs. (11) and this causes no problems when they have unique solution. Otherwise we would have to choose among the solutions and the natural choice is the one providing the global free energy minimum.

With $f_\alpha$ obtained from Eqs. (11) we can get full description of the spin ice thermodynamics in the BP approximation. Thus for the equilibrium values of spins, $\langle \sigma_\alpha \rangle$, we get using the lattice representation of Fig. 2b

$$\langle \sigma_\alpha \rangle = \frac{Tr_{\sigma_\alpha} \sigma_\alpha Z_\alpha^2(\sigma_\alpha) e^{h_\alpha \sigma_\alpha}}{Tr_{\sigma_\alpha} Z_\alpha^2(\sigma_\alpha) e^{h_\alpha \sigma_\alpha}} = \tanh(2f_\alpha - h_\alpha) \tag{13}$$

So according to Eq. 2 the equilibrium magnetization per spin is

$$\langle \mathbf{m} \rangle = \frac{1}{4}\sum_{\alpha=0}^{3} \mathbf{e}_\alpha \langle \sigma_\alpha \rangle. \tag{14}$$

Also from (8-11) we get the equilibrium entropy

$$S = -\frac{\partial F}{\partial T} = -\beta F - \frac{1}{4}\sum_{\alpha=0}^{3} h_\alpha \langle \sigma_\alpha \rangle + \frac{K}{Z(\mathbf{f})}\left(Z_1(\mathbf{f})e^{-2K} + 4Z_2(\mathbf{f})e^{-8K}\right) \tag{15}$$

Here we have introduced the quantities $Z_n(\mathbf{f})$, $n = 0,1,2$,

$$Z_n(\mathbf{f}) = \frac{1}{2}Tr_\sigma\left\{e^{\sum_{\alpha=0}^{3}\sigma_\alpha f_\alpha}\delta\left[\left(\sum_{\alpha=0}^{3}\sigma_\alpha\right)^2, 4n^2\right]\right\}, \tag{16}$$

which define the contributions to $Z(\mathbf{f})$ (8) from the groups of states with equal exchange energies

$$Z(\mathbf{f}) = \sum_{n=0}^{2} Z_n(\mathbf{f}) e^{-2n^2 K}. \qquad (17)$$

Thus $Z_0(\mathbf{f})$ corresponds to the contributions of spin ice states (two-in, two-out), $Z_1(\mathbf{f})e^{-2K}$ describes the contributions of (3-in, 1-out) and (3-out, 1-in) states while $Z_2(\mathbf{f})e^{-8K}$ results from all-in and all-out states.

Explicit expressions for $Z_n(\mathbf{f})$ read

$$Z_0(\mathbf{f}) = \cosh(f_0 + f_1 - f_2 - f_3) + 2\cosh(f_0 - f_1)\cosh(f_2 - f_3)$$

$$Z_1(\mathbf{f}) = \sum_{\alpha=0}^{3} \cosh\left(\sum_{\beta=0}^{3} f_\beta - 2f_\alpha\right) \qquad (18)$$

$$Z_2(\mathbf{f}) = \cosh\left(\sum_{\alpha=0}^{3} f_\alpha\right)$$

From (17), (18) we can also get the explicit form of the equations of state (11) as

$$\tanh(2f_\alpha - h_\alpha) = \frac{\partial \ln Z(\mathbf{f})}{\partial f_\alpha} \qquad (19)$$

Further differentiations of Eqs. (14), (15) could give us tensor of magnetic susceptibilities and specific heat.

Thus we have the quite simple theory of spin ice thermodynamics consisting essentially of Eqs. (8), (9). Yet such theory can describe all intricate features of this strongly frustrated system's behaviour in various fields as we show below. To start with one can easily find that in zero field $S$ (14) gives as the residual entropy at $T=0$ the Pauling value [19]

$$S_P = \frac{1}{2}\ln\frac{3}{2} \approx 0.2.$$

Indeed, at $H=0$ $f_\alpha = 0$ and $Z_0 = 3$, $Z_1 = 4$, $Z_0 = 1$ so

$$S(T) = -\frac{1}{2}\ln 2 + \frac{1}{2}\ln(3 + 4e^{-2K} + e^{-8K}) + 4K\frac{e^{-2K} + e^{-8K}}{3 + 4e^{-2K} + e^{-8K}}$$

Hence at $T=0$ ($K=\infty$) $S(0) = S_P$ while at $T=\infty$ ($K=0$) $S(\infty) = \ln 2$.
For the specific heat we obtain

$$C = \frac{24K^2 e^{-2K}\left(1 - 2e^{-2K} + 3e^{-4K}\right)}{\left(3 + e^{-2K} - e^{-4K} + e^{-6K}\right)^2}$$

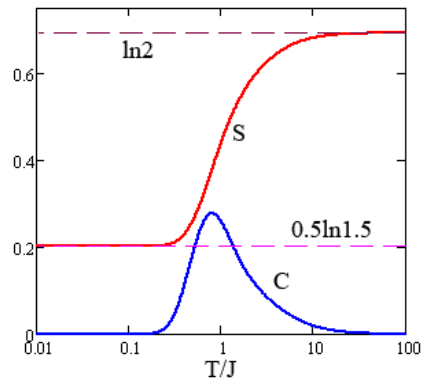

Fig.3. Temperature dependence of $S$ and $C$ in zero field. Note the logarithmic scale on the $T$ axis.

The temperature behaviour of S and C is shown in Fig.3. It illustrates the notions of "quasi-phases" and "pseudo-transition". S exhibits a broad crossover between Pauling and paramagnetic values yet below $T_m \approx 0.8J$ where C has a maximum we have the "spin-ice liquid" where local spin configurations are mainly "2-in, 2-out" ones while above $T_m$ they are almost uncorrelated and we have the "paramagnetic quasi- phase".

## 2. Spin-ice in [111] field.

For $\mathbf{H} = H\mathbf{e}_0$ we have
$h_0 = h, \ h_1 = h_2 = h_3 = -h/3, \ h \equiv \beta H$
and the solution to the Eq. (19) has apparently the form
$f_0 = x, \ f_1 = f_2 = f_3 = -y$
so it follows from (19)

$$\tanh(2x - h) = \frac{\partial \ln Z(x, y)}{\partial x} \tag{20}$$

$$3\tanh\left(2y - \frac{h}{3}\right) = -\sum_{i=1}^{3} \frac{\partial \ln Z(\mathbf{f})}{\partial f_i} = \frac{\partial \ln Z(x, y)}{\partial y}$$

where $Z(x, y)$ is given by Eq. (17) with

$$\begin{aligned}Z_0(x, y) &= 3\cosh(x + y) \\ Z_1(x, y) &= \cosh(x + 3y) + 3\cosh(x - y) \\ Z_2(x, y) &= \cosh(x - 3y)\end{aligned} \tag{21}$$

Also from (4), (13), (14) we have

$$\langle \mathbf{m} \rangle = \frac{\mathbf{e}_0}{4}\left[\tanh(2x - h) + \tanh\left(2y - \frac{h}{3}\right)\right] \tag{22}$$

Let us first consider the case of low temperatures and moderate fields
$T \ll J, \ H \leq J$.
As we show below here we can drop the contributions to Z proportional to $Z_1$ and $Z_2$ so equations (20) become

$$\tanh(2x - h) = 3\tanh\left(2y - \frac{h}{3}\right) = \tanh(x + y) \tag{23}$$

Hence we have
$x = y + h$,

$$m \equiv |\langle \mathbf{m} \rangle| = \tanh\left(2y - \frac{h}{3}\right) = \frac{t}{\sqrt{1 + 3t^2} + 1}, \quad t \equiv \tanh\left(\frac{4}{3}h\right) \tag{24}$$

Via this $m$ we can express other thermodynamic quantities

$$-\beta F = \frac{1}{2}\ln\frac{3}{2} + \frac{1}{8}\ln\frac{(1 - m^2)^3}{1 - 9m^2}$$

$$S = -\beta F - hm = \frac{1}{2}\ln\frac{3}{2} + \frac{3}{8}\left[(1 - m)\ln(1 - m) + (1 + m)\ln(1 + m)\right]$$

$$-\frac{1}{8}\left[(1 + 3m)\ln(1 + 3m) + (1 - 3m)\ln(1 - 3m)\right] \tag{25}$$

$$\chi' \equiv T\chi = \frac{\partial m}{\partial h} = \frac{2}{3}\frac{(1-m^2)(1-9m^2)}{1+3m^2}, \quad C = h^2\chi'$$

The relation for specific heat C follows from the equation

$$\frac{\partial \beta F}{\partial h} = -m \tag{26}$$

and the scaling form of $\beta F$ which depends only on $h = H/T$. Due to this scaling the above quantities are constant along the lines $H = cT$ and have different limiting values at $H = T = 0$ along these lines. In particular, at $H = 0$ we have

$$m = 0, \quad S = S_P, \quad \chi' = \frac{2}{3}, \quad C = 0, \tag{27}$$

while at $T = 0$

$$m = \frac{1}{3}, \quad S = \frac{1}{4}\ln\frac{4}{3} \approx 0.072, \quad \chi' = C = 0 \tag{28}$$

Here the reduced entropy with respect to $S_P$ is due to the partial lifting of the ground states' degeneracy in [111] field. It fixes the "out" direction for the 0 spin while three others can freely choose which one of them will also point out to obey the ice rule. This phase corresponds to "kagome ice" state in the original pyrochlore lattice [18].

Following [19] let us consider first the spins as belonging to $N/4$ independent tetrahedra so we get $\Gamma_0 = 3^{\frac{N}{4}}$ states for them. Then turning to the $N/4$ bond tetrahedra connecting the independent ones we see that their free (1, 2, 3) sites have $\sigma = 1$ with probability $p_+ = 1/3$ and $\sigma = -1$ with probability $p_- = 2/3$. Thus the average number of favourable 3-spin states (2-in, 1-out) per such tetrahedron is $3p_+ p_-^2 = \frac{4}{9}$ and total number of states $\Gamma = \left(\frac{4}{9}\right)^{\frac{N}{4}} \Gamma_0 = \left(\frac{4}{3}\right)^{\frac{N}{4}}$ gives the value of $S = N^{-1}\ln\Gamma$ from Eq.(28). So we see that Pauling-Anderson entropy estimates [19] are exact in the BP approximation.

The behaviour of thermodynamic variables near the point $H = T = 0$, given by Eqs. (25), is shown in Fig.4. The S and C temperature dependencies in the broader range $T \leq J$ for $H \ll J$ obtained via numerical solution of the equations of state are presented in Fig. 5. Note the appearance of additional low-T peak in $C(T)$ at $H \approx 0.7T$ designating the pseudo-transition between "kagome ice" and ordinary spin ice.

Now we can asses the range of validity of the above analytical results. We have assumed that

$$\varepsilon \equiv \frac{Z_1(x,y)e^{-2K} + Z_2(x,y)e^{-8K}}{Z_0(x,y)} \ll 1$$

This condition can be violated at $h \to \infty$ when $x, y \to \infty$. Indeed, in this case we have $\varepsilon \sim e^{2(y-K)}$ so the above results hold at $y < K$. From (24) we have $y \approx h/6$ at $h \to \infty$ and the validity condition for $T \ll J$ is $H < 6J$.

Thus for $T \ll J$ and $H \geq 6J$ we cannot ignore $Z_1, Z_2$. Yet analytical results in this case can be also readily obtained. Here $h \to \infty$ and we may assume that $x \to \infty$ so the second equation in (20) becomes

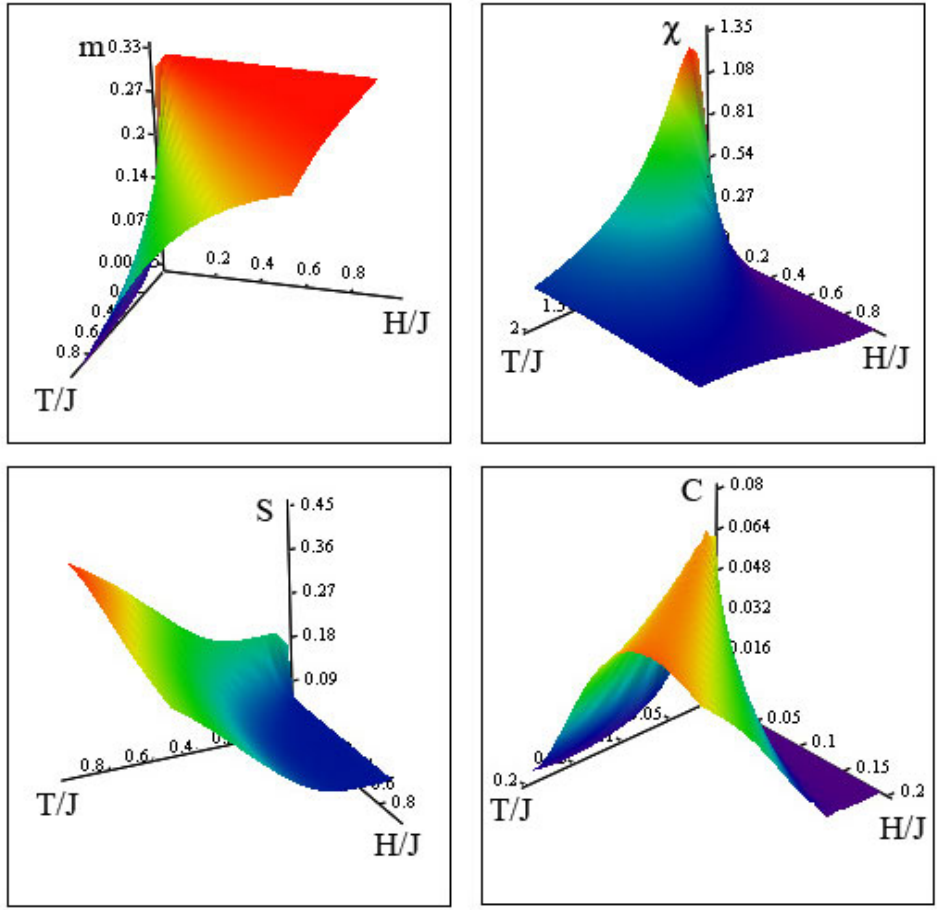

Fig.4. Field and temperature dependencies of thermodynamic variables near $H=T=0$ in [111] field.

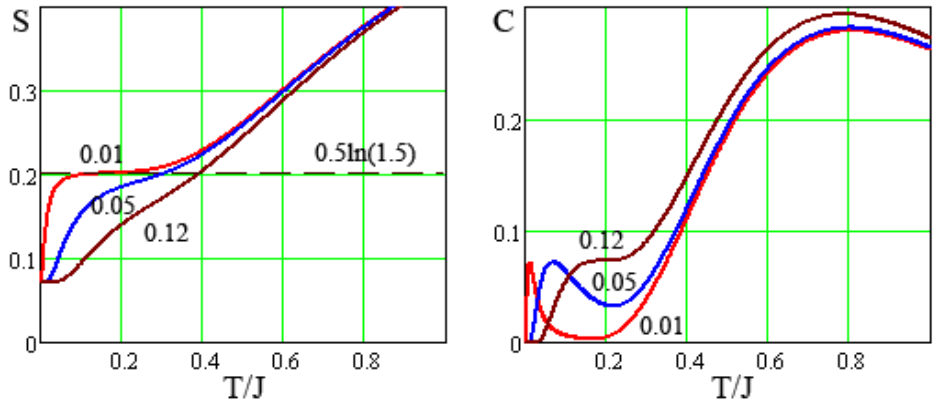

Fig.5. $S$ and $C$ temperature dependencies at $T \leq J$ for $H \ll J$, $\mathbf{H} \parallel [111]$. The values of $H/J$ are shown near the curves. Low-$T$ peak of $C$ defines the quasi-transition between "kagome ice" state with $S \approx 0.072$ to the spin ice with $S$ close to the Pauling entropy $S_P \approx 0.2$ (dashed line).

$$\tanh\left(2y - \frac{h}{3}\right) = \frac{1 + e^{2(y-K)}}{3 + e^{2(y-K)}}$$

This is actually the quadratic equation
$$u^2 - \lambda u - 2 = 0, \quad u \equiv e^{2y - h/3}, \quad \lambda \equiv e^{h/3 - 2K} \tag{29}$$
Thus
$$u \equiv e^{2y - h/3} = \frac{1}{2}\left(\lambda + \sqrt{\lambda^2 + 8}\right) \tag{30}$$

From the first equation in (20) we get the exact relation

$$e^{2x} = e^{2(y+h)} \frac{\varsigma(y,K)}{\varsigma(-y,K)}, \quad \varsigma(y,K) = 3 + e^{2(y-K)} + 3e^{-2(K+y)} + e^{-4(2K+y)} \quad (31)$$

Hence always $2x - h \to \infty$ contrary to $2y - h/3$ so according to (22)

$$m \equiv |\langle \mathbf{m} \rangle| = \frac{1}{4}\left[1 + \tanh\left(2y - \frac{h}{3}\right)\right] = \frac{\lambda + \sqrt{\lambda^2 + 8}}{\lambda + 3\sqrt{\lambda^2 + 8}} \quad (32)$$

Again using $2x - h \to \infty$ and Eqs. (30), (32) we can express thermodynamic quantities via magnetization

$$\begin{aligned}
-\beta F &= \frac{1}{2}\ln 2 + \frac{1}{4}\ln\frac{m^2}{1-2m} + \frac{h}{3} \\
S &= \frac{1}{2}(2-3m)\ln 2 + \frac{3}{2}m\ln m - \frac{3}{4}(1-2m)\ln(1-2m) \\
&\quad -(3m-1)\ln(3m-1) \\
\chi' &= \frac{\partial m}{\partial h} = \frac{2}{3}\frac{(3m-1)(1-2m)}{1-m}, \quad C = (h-6K)^2 \chi'
\end{aligned} \quad (33)$$

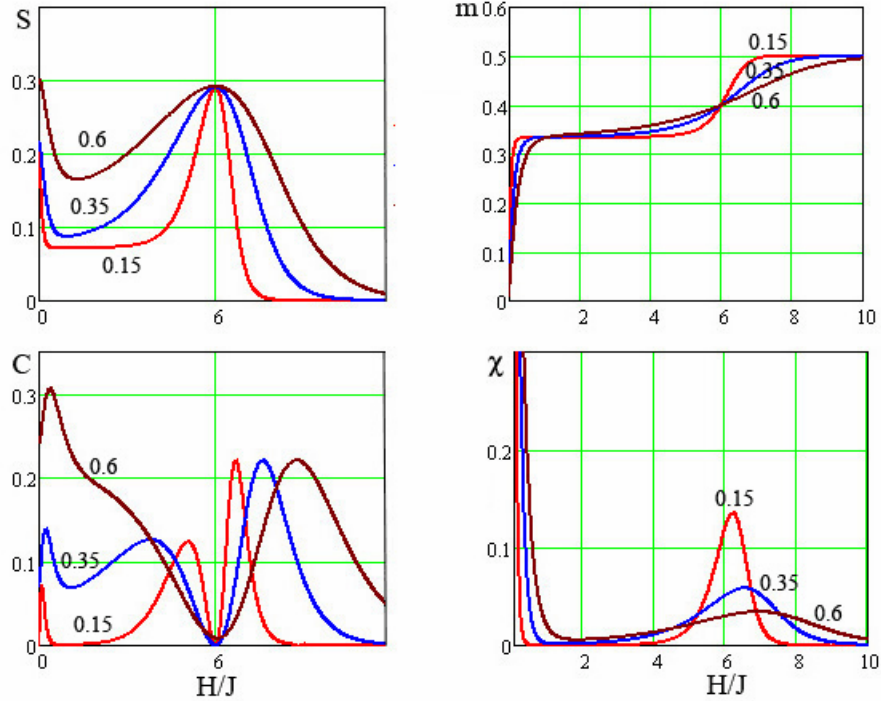

Fig.6. Field dependencies of thermodynamic variables, $T \ll J$, $\mathbf{H} \parallel [111]$. The values of $T/J$ are shown near the curves.

Thus at $T \ll J \le H$, the scaling dependence on the single parameter $(H - H_c)/T$, $H_c \equiv 6J$, also holds. Here thermodynamic quantities are constant along the straight lines emerging from the point $T = 0$, $H = H_c$. At this point spin ice undergoes the first order phase transition from the degenerate "kagome ice" phase to the completely ordered "3-in, 1-out" phase. At $T = 0$ we have

$$H < H_c, \quad m = \frac{1}{3}, \quad S = \frac{1}{4}\ln\frac{4}{3} \approx 0.072, \quad \chi = C = 0$$

$$H = H_c, \quad m = \frac{2}{5}, \quad S = \frac{1}{4}\ln\frac{16}{5} \approx 0.291, \quad \chi' = \frac{4}{225}, \quad C = 0 \qquad (34)$$

$$H > H_c, \quad m = \frac{1}{2}, \quad S = \chi = C = 0$$

The large value of $S$ at $H = H_c$ is due to the degeneracy between the configurations of adjoint phases. Note also that the results for $H < H_c$ coincide with the $T = 0$ results in Eq. (28) valid down to $H = 0$. So at $T = 0$, $0 < H < H_c$ we have plateaus in field dependencies of $S$ and $m$ which are also seen at low $T$, see Fig.6. Also they have the crossing points at $H = H_c$ due to scaling in agreement with experiments [6, 11] and simulations [18].

Fig. 7 shows the behaviour of thermodynamic variables at high $T$ and $H$. The location of specific heat ridges on the $H$-$T$ plane is summarized in Fig. 8. They are used to define the regions of spin ice, kagome ice, paramagnetic and completely ordered ("saturated") quasi-phases. Yet the attribution of some definite spin-liquid quasi-phase to the central region in Fig. 8 seems to be less appropriate as it is hard to describe the nature of collective excitations for the whole vast $H$ and $T$ ranges here.
The experimental data on $S$ and $C$ in [111] field [8-10] shows only vague resemblance to the present results. Accordingly the phase diagram in Fig. 8 differs essentially from the experimental ones [9, 11].

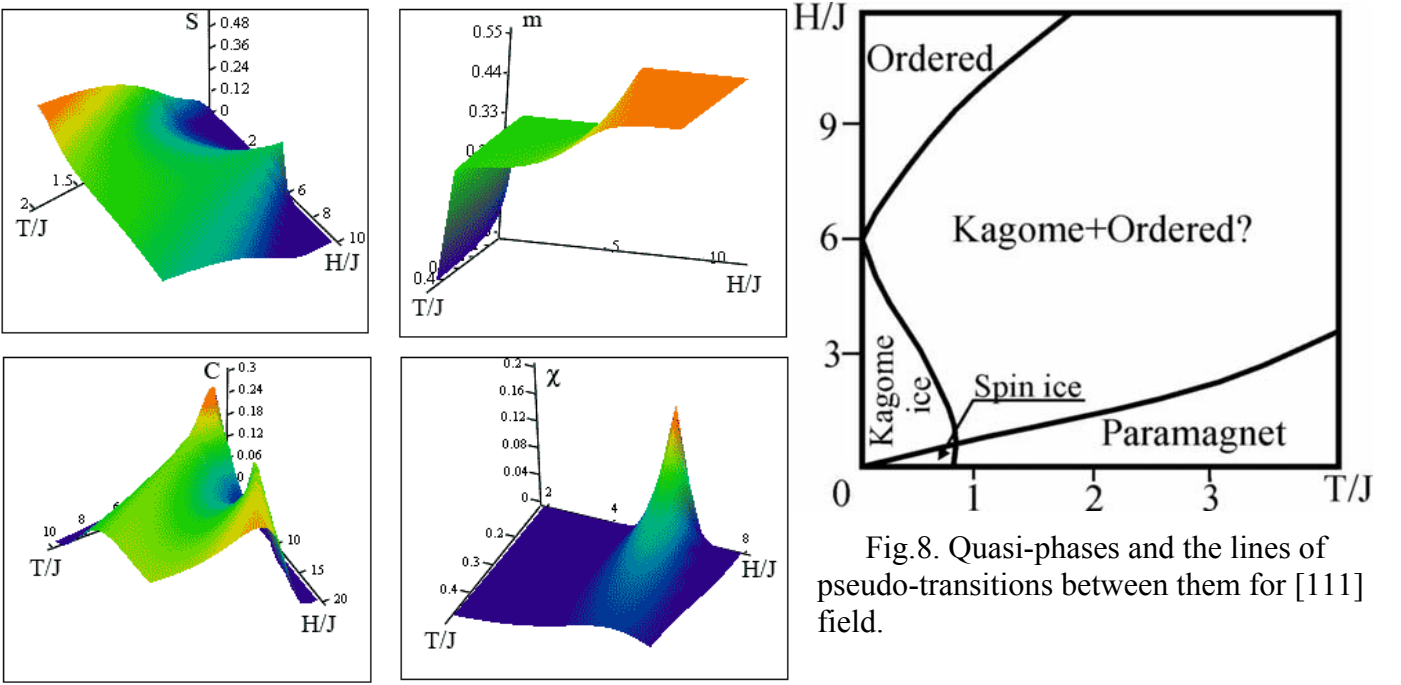

Fig.8. Quasi-phases and the lines of pseudo-transitions between them for [111] field.

Fig.7. Thermodynamic variables at high $T$ and $H$, $\mathbf{H} \parallel [111]$.

## 3. Spin-ice in [100] field.

For $\mathbf{H} = H\mathbf{e}_x$ we have

$$h_0 = h_1 = h/\sqrt{3}, \quad h_2 = h_3 = -h/\sqrt{3}, \quad h \equiv \beta H$$

so one of the six spin ice configurations (0, 1 spins out, 2, 3 spins in) becomes preferable in this field. Thus the ground state degeneracy is completely lifted and we would have $S=0$ at $T=0$. As at $H=0$ and low $T$ $S = S_P$ the growth of the field may result in the "pseudo-transition" from spin ice to the ordered "saturated" state. Indeed, it shows up in simulations [16, 17] and is extremely sharp at $T \ll J$ so one may define here the transition point [17]

$$H_c = T\frac{\sqrt{3}}{2}\ln 2 \approx 0.6T \qquad (35)$$

This sharpness results from the existence of specific string excitations with the unique free-energy gap which they can overcome at $H < H_c$ [17]. Yet here we show that this is only the "pseudo-transition" and its sharp anomalies can be described in BP approximation with the perfectly smooth functions.

The solution to the Eq. (19) has the form $f_0 = f_1 = -f_2 = -f_3 = x$ so

$$\langle \mathbf{m} \rangle = \frac{\mathbf{e}_x}{\sqrt{3}} \tanh\left(2x - \frac{h}{\sqrt{3}}\right) \qquad (36)$$

It follows from (17) - (19)

$$Z(x) = 2 + \cosh 4x + 4e^{-2K} \cosh 2x + e^{-8K}$$

$$Z(x)\tanh\left(2x - \frac{h}{\sqrt{3}}\right) = \sinh 4x + 2e^{-2K} \sinh 2x \qquad (37)$$

The last equation is actually the algebraic forth-order one for $v = \exp(-2x)$

$$av^4 + (2 + a^4 - b)v^3 + 3a(1-b)v^2 + (1 - 2b - a^4 b)v = ab \qquad (38)$$

$$a \equiv \exp(-2K), \quad b \equiv \exp\left(-\frac{2}{\sqrt{3}}h\right)$$

It has a unique solution in the interval $0 < v < 1$ which depends analytically on $H$ and $T$ at $T > 0$. Finding it numerically we get the thermodynamic quantities as functions of $H$ and $T$. The results are shown in Figs. 9, 10. Quite remarkably we see at low $T \ll J$ absolutely sharp anomalies which can be easily mixed up with the genuine phase transitions. But in present theory they are only sharp crossovers, the genuine first-order transition take place only at $T = H = 0$.

To show this let us consider low $T \ll J$ and determine the behaviour of $v$ when $a \to 0$. Here we can simplify Eq. (38) to

$$(2-b)v^3 + (1-2b)v = ab \qquad (39)$$

For $a \to 0$ we have

$$b < \frac{1}{2}, \quad v \to 0; \quad b = \frac{1}{2}, \quad v = \left(\frac{a}{3}\right)^{1/3}; \quad b > \frac{1}{2}, \quad v \to \sqrt{\frac{2b-1}{2-b}}.$$

Thus $v = v(b)$ at small $a \to 0$ is very close to its limiting function

$$v = \sqrt{\frac{2b-1}{2-b}} \vartheta(2b-1) \qquad (40)$$

differing from it by the small amount of the order $a^{1/3} = e^{-2K/3}$.

So we can describe the low-temperature thermodynamics very precisely with the singular $v = v(b)$ from Eq. (40). Its singular point

$$b \equiv e^{-\frac{2H}{\sqrt{3}T}} = \frac{1}{2}$$

coincides with the critical field $H_c$ (35) below which the string fluctuations develop in the original pyrochlore structure [17].

Thus we have at $H > H_c$ ($b < 1/2$)

$$x = \infty, \quad m = \frac{1}{\sqrt{3}}, \quad S = C = \chi = 0$$

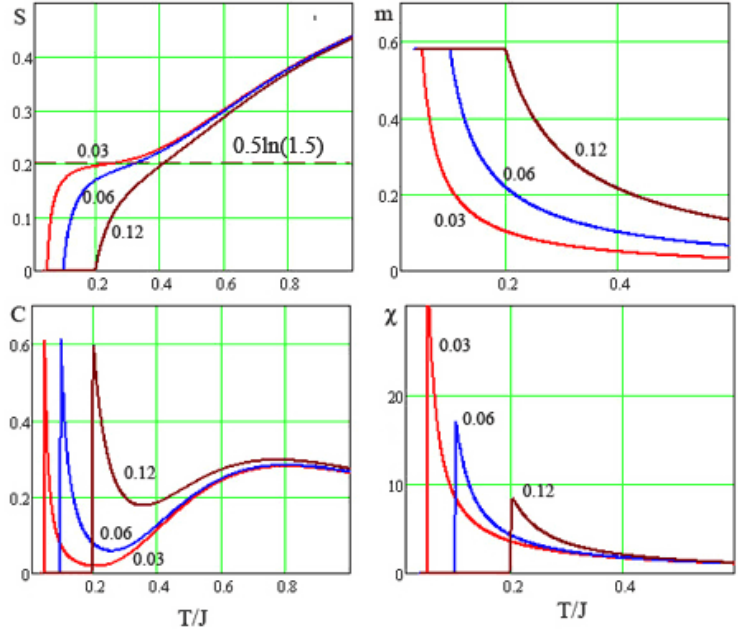

Fig.9. Temperature dependencies of thermodynamic variables in low [100] field. The values of $H/J$ are shown near the curves. Peaks of $C$ and $\chi$ define the Kasteleyn pseudo-transition from the ordered quasi-phase to the spin ice with $S$ close to $S_P = 0.5\ln(1.5)$ (dashed line).

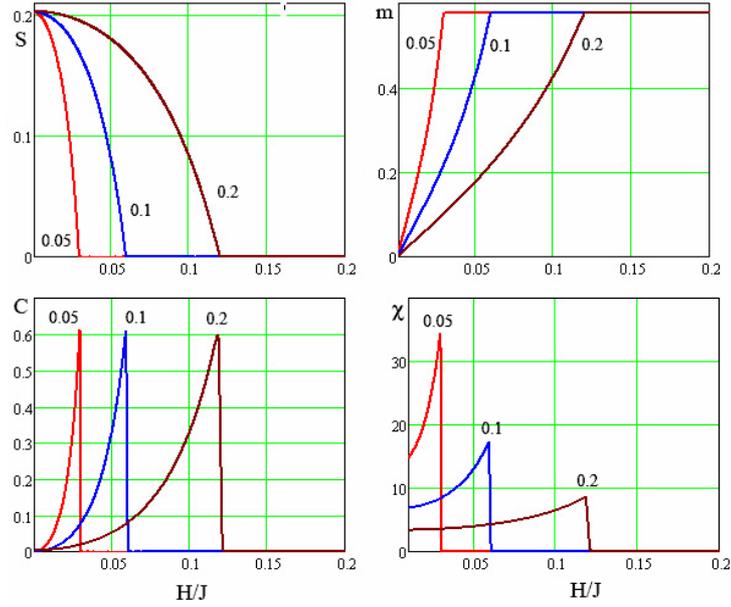

Fig.10. Field dependencies of thermodynamic variables in [100] field at $T \ll J$. The values of $T/J$ are shown near the curves.

while at $H < H_c$ ($b > 1/2$) we get

$$-\beta F = \frac{1}{2}\ln\frac{3b}{4b - b^2 - 1},$$

$$S = -\beta F - hm = \frac{1}{2}\ln\frac{3}{4b - b^2 - 1} - \frac{2h}{\sqrt{3}}\frac{b(2-b)}{4b - b^2 - 1},$$

$$m = \frac{1}{\sqrt{3}}\frac{1 - b^2}{4b - b^2 - 1},$$

$$\chi' \equiv T\chi = \frac{8}{3}b\frac{1-b+b^2}{(4b-b^2-1)^2}, \quad C = h^2\chi'.$$

At $H = H_c - 0$ ($b = 1/2 + 0$)

$$\chi' = \frac{16}{5}, \quad C = \frac{4}{3}\ln^2 2, \quad m = \frac{1}{\sqrt{3}}, \quad S = 0$$

These expressions agree quite well with the numerical data for $T \leq 0.2J$ shown in Figs. 9, 10 and with the results of Monte-Carlo simulations [16, 17].

Fig. 11 presents the general view of $S$ and $C$ behaviour over $H$-$T$ plane and Fig.12 depicts the location of the specific heat ridges defining the quasi-transitions between paramagnetic, spin ice and saturated quasi-phases. Note that dashed line in Fig. 12 does not correspond to the local maximum of $C$ as it vanishes at higher fields, it is only continuation of the low-field ridge. So the exact determination of "quasi-phases" from $S$ and $C$ behaviour is not always possible.

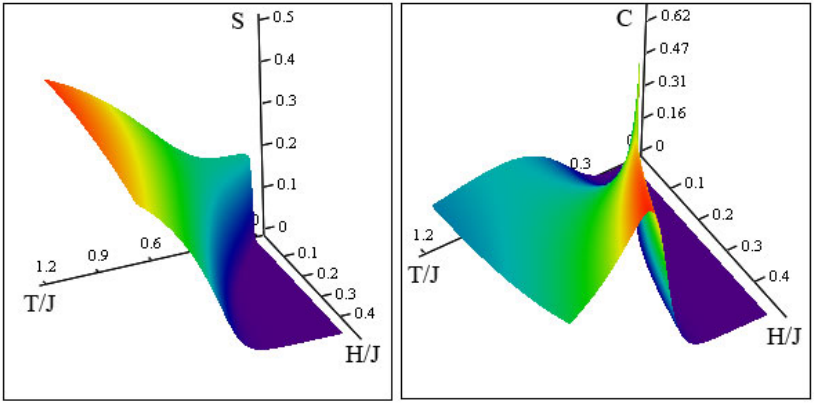
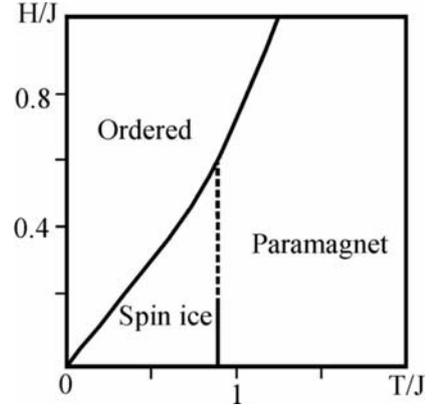

Fig.11. $S$ and $C$ in [100] field.     Fig.12. Pseudo-phases in [100] field. Dashed line continues the low-field ridge.

The existing experiments in [100] field [6, 8, 12] give only scarce qualitative evidences in favour of present results far from $T = H = 0$ point.

## 4. Spin-ice in [011] field.

For $\mathbf{H} = \frac{H}{\sqrt{2}}(\mathbf{e}_y + \mathbf{e}_z)$ we have

$$h_0 = -h_1 = \sqrt{\frac{2}{3}}h, \quad h_2 = h_3 = 0, \quad h \equiv \beta H.$$

Thus the field in [011] direction does not act on 2, 3 spins which form the perpendicular to the field $\beta$-chains while 0, 1 spins belong to the $\alpha$-chains oriented along the field [13]. At $T=0$ the directions of spins in $\alpha$-chains are fixed by the field (0-out, 1-in) but the $\beta$-chains' spins have two possibility to comply with the spin ice rule and provide the lowest energy – either all 2 spins are fixed at -1 and all 3 spins at +1 or vice versa. Thus for the $\beta$-chains of the length $L$ we have $\Gamma = 2^{\frac{N}{2L}}$ ground state configurations. Yet in spite of the macroscopic degeneracy the zero temperature entropy

$$S = \frac{1}{N}\ln\Gamma = \frac{1}{2L}\ln 2$$

tends to zero in the thermodynamic limit. Hence here we may expect the existence of crossover between spin ice and "ordered chains" quasi-phases with the growth of the field at low $T$.

The thermodynamics in such field is described analytically in rather simple way. The solution to the equation of state has the form

$f_0 = -f_1 = x$, $f_2 = f_3 = 0$

so

$$\langle \mathbf{m} \rangle = \frac{1}{2\sqrt{3}} \tanh\left(2x - \sqrt{\frac{2}{3}}h\right) \frac{\mathbf{H}}{H}$$

$$Z(x) = 2(1 + e^{-2K})[\cosh(2x) + d(K)], \quad d(K) = \frac{1 + 2e^{-2K} + e^{-8K}}{2(1 + e^{-2K})}$$

and equation of state

$$\tanh\left(2x - \sqrt{\frac{2}{3}}h\right)[\cosh(2x) + d(K)] = \sinh(2x)$$

can be easily solved to give

$$m \equiv |\langle \mathbf{m} \rangle| = \frac{1}{\sqrt{6}} \frac{\sinh\left(\sqrt{\frac{2}{3}}h\right)}{\sqrt{\sinh\left(\sqrt{\frac{2}{3}}h\right)^2 + d(K)^2}} \qquad (41)$$

$$-2\beta F = \ln(1 + e^{-2K}) + \ln\left[\cosh\left(\sqrt{\frac{2}{3}}h\right) + \sqrt{\sinh\left(\sqrt{\frac{2}{3}}h\right)^2 + d(K)^2}\right] \qquad (42)$$

We do not present here the general expressions for $S$, $C$ and $\chi$ as they are rather cumbersome. Yet at $T \ll J$ $d(K) \approx 1/2$ and we get the following scaling functions of $h$

$$m = \sqrt{\frac{2}{3}} \frac{t}{\sqrt{1 + 3t^2}}, \quad t \equiv \tanh\left(\sqrt{\frac{2}{3}}h\right)$$

$$-\beta F = \frac{1}{2}\ln\frac{3}{2} + \frac{1}{2}\ln\frac{\sqrt{1-t^2}}{2 - \sqrt{1+3t^2}}, \quad S = -\beta F - hm$$

$$\chi' \equiv \frac{\partial m}{\partial h} = \frac{2}{3} \frac{1-t^2}{(1+3t^2)^{3/2}}, \quad C = h^2 \chi'.$$

Thus at $H = 0$ $S = S_P$, $C = m = 0$, $\chi' = \frac{2}{3}$

while at $T = 0$ $S = C = \chi' = 0$, $m = \frac{1}{\sqrt{6}} \approx 0.41$.

Accordingly the specific heat $C(T)$ shows a maximum at $T \ll J$, $H_c \approx 0.86T$ indicating the crossover between spin ice and "ordered chains" quasi-phases in addition to the one at $T \approx 0.8J$ for the paramagnet-spin ice quasi-transition, cf. Fig (13).

The other specific quasi-phase exists in [011] field at $J \ll T \ll H$. Indeed, here we have from (41), (42)

$$m = \frac{1}{2\sqrt{3}}, \quad -\beta F = \frac{1}{2}\ln 2 + \frac{h}{\sqrt{6}}, \quad S = \frac{1}{2}\ln 2 \approx 0.347.$$

In this region strong field fixes the directions of 0, 1 spins but $T$ is sufficiently high to initiate free flips of spins in $\beta$-chains. These flips has no effect on $m$ determined solely by 0, 1 spins but give rise to the one half of the paramagnetic entropy as the half of the spins in a system fluctuate between two states. As a result we have two maxima of $C(T)$ – one at $T \approx 0.8J$ indicating the crossover between "random $\beta$-chains" and "ordered chains" quasi-phases and another at $T \approx 0.9H$ corresponding to the "random $\beta$-chains" – paramagnet quasi-transition, see Fig. (14).

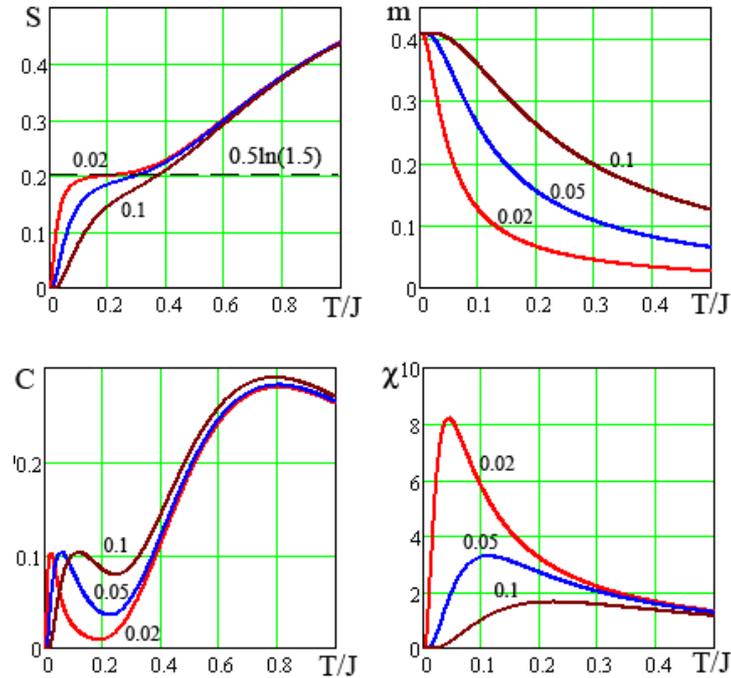

Fig.13. Temperature dependencies of thermodynamic variables in low [011] fields. The values of $H/J$ are shown near the curves. $S$ shows the approach to the Pauling value $S_P$ = 0.5ln1.5 (dashed line) and low-$T$ peak of $C$ defines the pseudo-transition from spin ice to ordered chains.

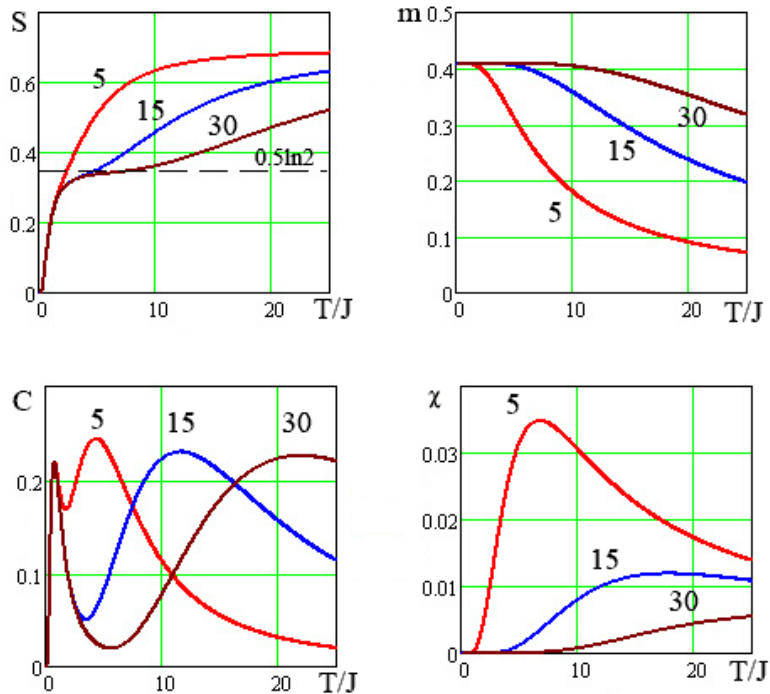

Fig.14. Temperature dependencies of thermodynamic variables in high [011] fields. The values of H/J are shown near the curves. The high-field $S$ shows the approach to the "random β-chains" value 0.5ln2≈0.347 (dashed line) in the region delimited by $C$ maxima.

Overall picture of the $C(T, H)$ relief is shown in Fig. (15). Its ridges are used to define the boundaries between quasi-phases presented in Fig. (16). Again we have no first-order transitions anticipated on the basis of simulations [12]. Experimental $C$ in [011] field and resulting phase diagram [13] agree qualitatively with Figs (15, 16) except for the close vicinity of $H=T=0$ point.

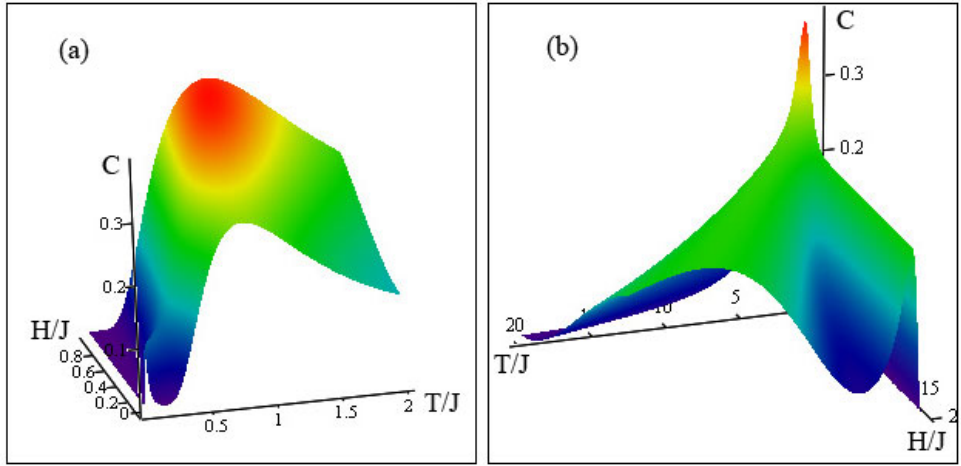
Fig.15. Specific heat in low (a) and high (b) [011] fields.

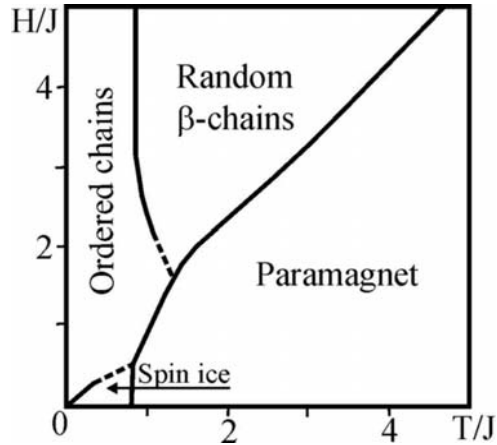
Fig.16. Pseudo-phases in [011] field. Dashed lines are the continuation of the vanishing ridges of *C*.

## 5. Conclusions

The main conclusion from the present results amounts to the recognition of BP approximation's ability to adequately describe many intricate features of thermodynamics of short-range frustrated magnet in a field. In some way it is superior to the Monte-Carlo simulations as it helps to discern the true phase transitions and the sharp crossovers. Thus in present model in cases considered the true first-order transitions take place only at $T = H = 0$ and at $T = 0$, $H = 6J$ points while the sharp field-induced anomalies result from their proximity. It is quite possible that this feature will be preserved in the rigorous theory of short-range 3d spin ice as present results resemble closely numerically exact Monte-Carlo data. More generally we may note that BP approximation describes adequately the short-range correlations and neglects the long-range ones and this is sufficient to give a qualitatively correct picture of thermodynamics near first-order transitions where correlation length stays finite. Evidently, it would be less successful in the vicinity of the second-order transitions.

We should also note that the BP approximation is a variant of more general cluster variation methods [20, 21] which can give adequate quantitative results for the systems with strong short-range competing interactions (contrary to the simple mean-field approximation). In particular the application of cluster methods to the crystals with tetrahedral units in their structures such as KDP-type ferroelectrics [22] and ordering alloys [21, 23] have also given the results which can be compared favorably with those of Monte-Carlo simulations. So we may suppose that further development of cluster approximation beyond the BP one along the lines presented in Refs. [20, 21] could further improve the description of the thermodynamics of the spin ice compounds.

Another point demonstrated by the BP results is the usefulness of the notions of "quasi-phases" and "pseudo-transitions". For example, they relieve us from the necessity to guess what is the order parameter for the Kasteleyn transition or is it of the first- or of the second-order type? Also they allow us to discern the definite spin-liquid states even at very high temperatures and fields which is the case of the "random β-chains" quasi-phase in [011] field, see Fig. 16. Yet it is not always possible to assign the definite type of spin excitations to the region surrounded by the specific heat ridges, cf. Fig. (8), and these ridges may vanish to leave the boundary between quasi-phases indefinite, see Figs. (12, 16).

Also the present results reproduce nearly quantitatively the experimental data for the entropy and magnetization of spin ices in fields. Yet experimental specific heats in the vicinity of the mentioned zero-temperature transitions show much less agreement with theory. The origin of these discrepancies still has to be found in future studies.

Author gratefully acknowledges the useful discussions of this work with S.E. Korshunov.